\documentclass{moriond}

\usepackage[utf8]{inputenc} 

\def\be{\begin{equation}}
\def\ee{\end{equation}}
\def\bea{\begin{eqnarray}}
\def\eea{\end{eqnarray}}

\newcommand{\Photo}{\includegraphics[height=35mm]{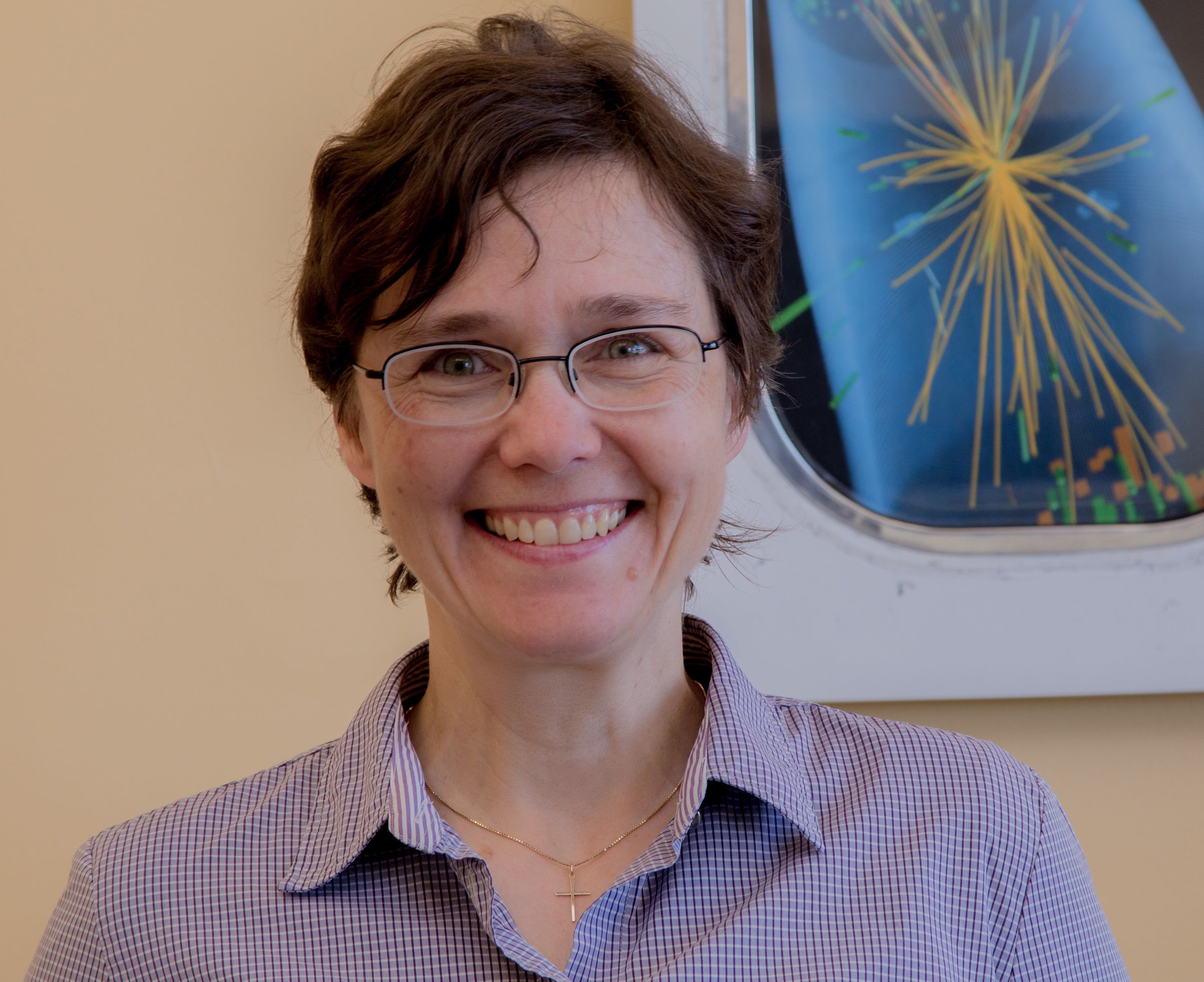}}

\begin{document}
\vspace*{4cm}
\title{QCD AND HIGH ENERGY INTERACTIONS: MORIOND 2019 THEORY SUMMARY}

\author{ D. WACKEROTH }

\address{Department of Physics, University at Buffalo, 239 Fronczak Hall,\\
Buffalo, NY 14221, U.S.A.}

\maketitle\abstracts{Highlights of recent theory developments are
  summarized relevant to precision Standard Model (SM) studies and
  searches for Beyond-the-SM (BSM) phenomena at present and future
  high-energy $pp$ and $e^+ e^-$ colliders, and $B$-factories, as well
  as to selected topics in heavy ion collisions.}

\section{Introduction}

We live in exciting times where a wealth of data from a wide range of
experiments (see, e.g., the experimental summary by
V.~Vagnoni~\cite{Vagnoni:2019muz}) allows us to probe all aspects of
the SM and the computational framework of Quantum Field Theory, often at an
unprecedented level of precision, and to perform new and increasingly
sensitive searches for BSM physics. At this conference, we were
treated to an impressive line-up of talks on recent theory
developments in a wide range of topics. In the following, I will
provide a brief summary of the presented results and studies and I
refer to the corresponding publications and contributions to these
proceedings for more details.

\section{BSM searches in flavor-changing processes in $B$ meson decays}

Flavor-changing processes in the quark and lepton sectors provide an
indirect window to BSM physics, which have the potential to probe high
energy scales of new physics (NP) complementary to direct searches for
new particles at the LHC. Several measurements of flavor observables
in $B$ meson decays by the ATLAS, CMS, LHCb, Belle and BARBAR show tensions with
the SM predictions (see, e.g., Ref.~\cite{Vagnoni:2019muz} for a recent
overview). While it can well be that these {\em flavor anomalies} are
statistical fluctuations, underestimated experimental systematics or
theoretical uncertainties, it is still interesting to confront them
with specific NP scenarios or in a model-independent effective field
theory (EFT) approach, to see whether a consistent picture
emerges. Before presenting the highlights of some of these studies,
let's first discuss an example of how theory uncertainties can be
further reduced, for instance by new ideas for a precision,
model-independent extraction of the CKM mixing matrix element $V_{cb}$
from data presented in~\cite{Fael:2018vsp}. $V_{cb}$ is not only an
important input parameter for heavy flavor observables but also a
sensitive probe of NP, for example NP may not obey the SM CKM unitarity
relation. An extraction of $V_{cb}$ from inclusive $B \to X_c l \bar
\nu$ decays relies on Heavy Quark Expansion (HQE) (see, e.~g., Ref.~\cite{Manohar:2000dt} for a review),
which allows the moments of kinematic distributions to be written as a
series in $\alpha_s$ and $\Lambda_{QCD}/m_b$. The series involves
non-perturbative HQE parameters, which need to be extracted from
data. However, the higher the order in $\alpha_s$ and $1/m_b$, the
higher the number of HQE parameters, e.~g., at ${\cal O}(1/m_b^4)$
there are 9 and 13 parameters at tree-level and ${\cal O}(\alpha_s)$, respectively. The
current method can handle up to ${\cal O}(1/m_b^3)$ but for further
improvements in the theory uncertainty ${\cal O}(1/m_b^4)$ should be included
as well. In making use of a known reparametrization invariance, which
links the HQE parameters at different orders and thus can reduce the
number of independent parameters, an alternative model-independent
extraction of $V_{cb}$ from data is proposed
in~\cite{Fael:2018vsp}. This promising approach based on using the
moments of the leptonic invariant mass spectrum can be already
tested with existing BARBAR and Belle data.

Among the flavor anomalies the ones observed in the precisely measured
(by LHCb~\cite{Aaij:2019wad}) lepton flavor universal (LFU) ratios $R_M$ of
flavor changing neutral current (FCNC) processes $b\to s l^+l^-$ ($M=K,K^*$ and
$q^2=m_{ll}^2, l=e,\mu$):
\begin{equation}\label{eq:DWone}
R_M[q_{min}^2;q_{max}^2]=\frac{\int_{q_{min}^2}^{q_{max}^2} dq^2 d\Gamma(B \to M \mu^+ \mu^-)/dq^2}{\int_{q_{min}^2}^{q_{max}^2} dq^2 d\Gamma(B \to M e^+ e^-)/dq^2}
\end{equation} 
are especially interesting probes of NP: the theoretical uncertainties
are under such good control that a deviation from unity larger than
about 1\%~\cite{Bordone:2016gaq} could be interpreted as a signal of
LFU violating (LFUV) NP. In~\cite{Alguero:2019pjc}, the impact of
$R_M$ together with other anomalies observed in $b \to sl^+l^-$
transitions was studied by performing a global fit in a
model-independent approach based on the effective
Hamiltonian~\cite{Grinstein:1987vj,Buchalla:1995vs}:
\begin{equation}\label{eq:DWtwo}
{\cal H}_{e\!f\!f}(b\to s\gamma^*) =- \frac{4 G_F}{\sqrt{2}} V_{ts}^* V_{tb} \sum_i C_i {\cal O}_i 
\end{equation} 
Here the heavy degrees of freedom above the electroweak scale
($t,H,W,Z$ and possible NP) have been integrated out in short-distance
Wilson coefficients $C_i$.  NP can either modify the ten main SM
Wilson coefficients or introduce additional operators. In the global
fit to $b\to sl^+l^-$ data of~\cite{Alguero:2019ptt} many different NP
scenarios considering both LFU NP and LFU violating (LFUV) NP have
been found to be in good agreement with the data. With more precise
measurements these EFT results can serve as guidance for the
construction of specific NP models.

Naturally the NP scenarios consistent with $B$ meson decay observables
also need to be confronted with other flavor observables or
electroweak precision observables, which can consistently be done in a
model-independent way by performing a global fit in the SM EFT (SMEFT)
approach~\cite{Buchmuller:1985jz} (and proper matching to the
aforementioned low-energy EFT valid at scales smaller than the
electroweak scale).  SMEFT assumes that the UV-complete NP model is
beyond the reach of direct observation and thus only manifests itself
in form of higher-dimensional operators built from SM fields (it is
also assumed that the SM gauge symmetry is preserved). For a recent
SMEFT global fit to flavor data see, e.~g.,
Ref.~\cite{Aebischer:2019mlg}. Within SMEFT, NP contributions in the
extraction of SM input parameters can be consistently taken into
account, as discussed in~\cite{Descotes-Genon:2018foz} on the example
of CKM mixing matrix elements $V_{ij}$ extracted from a suggested
subset of four flavor observables.  The proposed procedure
in~\cite{Descotes-Genon:2018foz} allows for a separation of NP effects
originating from the extraction of $V_{ij}$ and those affecting the
flavor observables included in the global fit.

Examples of specific NP models, which already contribute at tree-level
and are consistent with $B$ anomalies, are models with an extra $Z'$
boson, Lepto-Quarks (LQ), and a charged Higgs boson.  A specific SM
extension consistent with $R_M$ and which has the added benefit that
it could explain the hierarchy of fermion masses and mixing, is a
flavor-dependent, spontaneously broken, anomaly-free $U(1)_F$
extension of the SM studied
in~\cite{Davighi:2019jwf}. There, it is assumed
that the heavy $U(1)_F$ gauge boson $Z'$ only couples to the third
family (which is also motivated by the fact that there are no
deviations from SM predictions in the decay of lighter mesons), where
its couplings are fixed by gauge anomaly cancellation conditions. In
this model, $R_M$ is affected via tree-level exchange of a $Z'$ with
couplings to $b_L \bar s_L, \mu_L^+ \mu_L^-$ due to mixing with the SM $Z$
boson. Among the considered observables apart from $R_M$, are LFU
tests at LEP, direct searches for the $Z'$ boson at the LHC, and
$B_s-\bar B_s$ mixing, and interesting bounds on the parameters space
of the model are extracted as well as the prospects for a full
coverage at the HL-LHC are discussed~\cite{Allanach:2019iiy}.

A minimal $Z'$ model where the $Z'$ does not have significant
couplings to the muon and the implications for the model parameters
$C_{1q}, q=u,d$ from both $B$ anomalies (see also, e.g.,
\cite{DAmico:2017mtc}) and the weak charge of the Cesium $Q_W^{CS}$
atom and the proton $Q_{W}^p$ has been considered
in~\cite{DAmbrosio:2019tph}.  Interestingly, there is only a small
overlap between bounds from $B$ anomalies and $Q_W^{CS,p}$ and
including the former can provide additional discriminating information.

\begin{figure}
\begin{minipage}{0.66\linewidth}
\centerline{\includegraphics[width=0.7\linewidth]{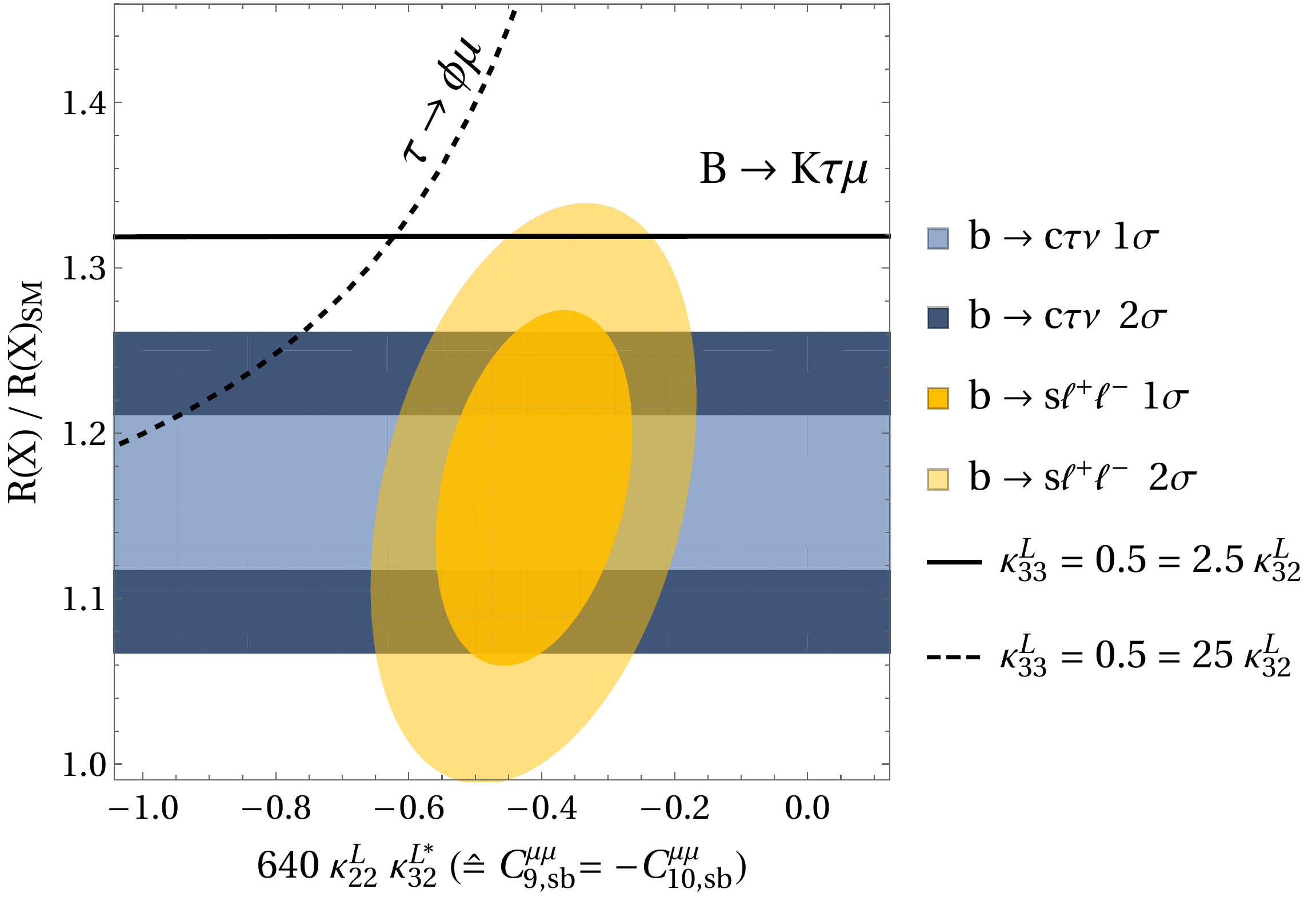}}
\end{minipage}
\caption[]{Allowed (colored) regions in the
  $R(X)/R(X)_{SM}-\kappa_{22}^L \kappa_{32}^{L*}$ plane ($X=D,D*$) at
  the $1 \sigma$ and $2 \sigma$ level. Taken from
  \protect\cite{Crivellin:2019szf}.}
\label{fig:DWone}
\end{figure}

Another intriguing NP explanation for the observed $B$ anomalies is a
vector LQ $SU(2)_L$ singlet ($V_\mu^1$) with hypercharge $-4/3$
arising in the Pati-Salam model~\cite{Pati:1974yy}, which affects both
the charged current $b \to c\tau \nu_\tau$ and FCNC $b\to s \mu \mu$
transitions via tree-level exchange of a vector
LQ~\cite{Crivellin:2018yvo,Crivellin:2019szf}. As discussed
in~\cite{Crivellin:2018yvo}, in this model there are only
loop-suppressed effects in flavor observables, which agree with SM
predictions such as $b\to s\gamma$, but one can strongly enhance $b\to
s \tau \tau$ transitions and can induce large loop effects in $b\to s
\mu \mu$. Figure~\ref{fig:DWone} shows the allowed parameter space of
the couplings $\kappa_{fi}^L$ of the LQ to SM particles described by
the Lagrangian ${\cal L}=\kappa_{fi}^L \bar Q_{f} \gamma^\mu L_i
V_\mu^{1\dagger}$~\cite{Crivellin:2019szf}.

The $b\to c\tau \nu$ transitions can also be affected by tree-level
exchanges of a scalar LQ or a charged Higgs boson. Combinations of
Wilson coefficients corresponding to these scenarios have been
considered in fits to tauonic $B$ decay observables
in~\cite{Blanke:2018yud,Blanke:2019qrx}. While predictions for
individual decay rates come with large theoretical uncertainties due
to their dependence on hadronic form factors and $V_{ij}$ parameters,
in the branching fraction ratios $R(D^{(*)})={\rm BR}(B \to
D^{(*)}\tau \nu)/{\rm BR}(B\to D^{(*)} l\nu)$ the $V_{ij}$ cancel and
the uncertainties originating from the form factors are reduced. They
are thus sensitive probes of NP and the inclusion of the tau
polarization asymmetry $P_\tau(D^*)$ measured by Belle could help to
distinguish between different NP scenarios~\cite{Blanke:2018yud}.
Apart from theoretically well controlled observables such as ratios
$R_M$, very rare $B$ meson decays, which are strongly suppressed in
the SM, are also ideal for the search for indirect signals of NP. An
example is the doubly weak transition $b \to d d \bar s$ due to a box
diagram in the SM, studied in~\cite{Bhutta:2019rdf} in the exclusive
wrong-sign weak decay $\bar B^0 \to K^+ \pi^-$.  An enhancement of up
to six orders of magnitude over the SM prediction for the decay rate
is found in two variants of the Randall-Sundrum model for a wide range
of model parameters, which could push future NP searches in this decay
at Belle-II and LHCb into the range of observability~\cite{Bhutta:2019rdf}.

\section{BSM searches at the LHC beyond simplified models}

Direct searches for SUSY particles and their interpretation in terms
of bounds on the SUSY parameter space at the LHC may rely on
phenomenological versions of the minimal supersymmetric SM (MSSM) with
a reduced set of parameters or more often on so-called simplified
models (see, e.g., the review {\em SUSY: experiment} in
\cite{Tanabashi:2018oca}). Exploring LHC data beyond these
simplifications is becoming increasingly important and may reveal
something unexpected.  For instance, an interpretation of LHC searches
in the full neutralino and chargino sector of the MSSM has been
performed within the {\sc GAMBIT} framework
in~\cite{Athron:2017yua,Kvellestad:2019fdw} and it turns out that a
light SUSY scenario is preferred by the global fit with the masses of
the lightest(heaviest) bino-like neutralino to be about 200(700)
GeV. Furthermore, non-minimal realizations of SUSY as the one studied
in~\cite{Chalons:2018gez} can have very distinct signatures which have
not yet been probed by LHC searches.  Ref.~\cite{Chalons:2018gez}
investigates the phenomenological consequences of the Minimal Dirac
Gaugino MSSM (MDMSSM), which allows for Dirac masses for gauginos, and
after SUSY breaking contains a scalar and pseudo-scalar sgluon.  An
interesting feature of the MDMSSM is that gluino(squark)-pair
production cross sections are significantly enhanced(reduced) compared
to the MSSM. The impact on existing bounds on squark or gluino masses
can be studied by recasting LHC analyses done in the context of
simplified model, and significant effects have been found in a number
of benchmark scenarios~\cite{Chalons:2018gez}. Another non-minimal
extension of SUSY is studied in~\cite{DelleRose:2018rpp}, where the
MSSM is extended in such a way that a right-handed sneutrino emerges
as a viable dark matter candidate, either by adding a gauged $(B-L)$
symmetry (BLSSM) ~\cite{DelleRose:2017uas}, or by supersymmetrizing the
SM extended by three heavy neutrinos.  These models have the
attractive feature that they address the origin of both DM and
neutrino masses. Again existing searches can be recast to obtain
bounds on the parameter space of these models as discussed
in~\cite{DelleRose:2018rpp}, where it has been also found that
sneutrino DM can be better accommodated by relic density limits than
MSSM neutralino DM.

\section{Precision calculations for SM and BSM studies at the LHC}

\begin{figure}
\begin{minipage}{0.44\linewidth}
\centerline{\includegraphics[width=0.8\linewidth]{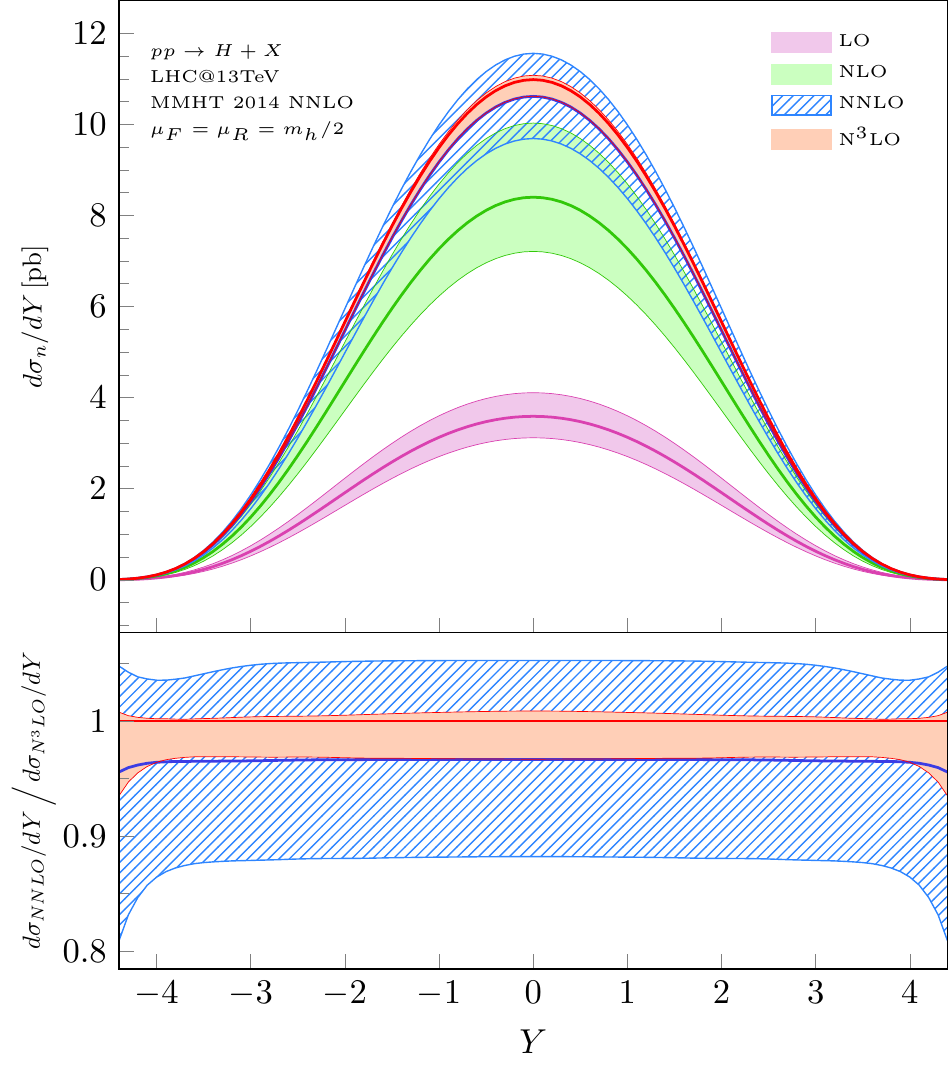}}
\end{minipage}
\hfill
\begin{minipage}{0.43\linewidth}
\centerline{\includegraphics[width=0.8\linewidth]{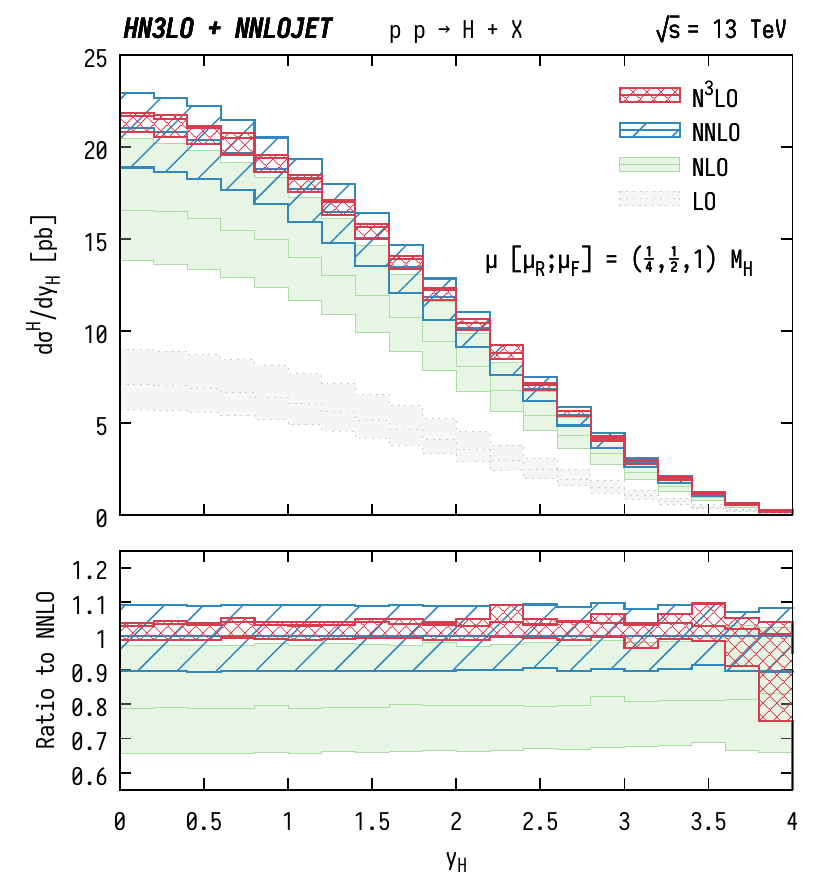}}
\end{minipage}
\caption[]{Predictions for the SM Higgs boson rapidity distribution 
at LO, NLO, NNLO and N$^3$LO at the 13 TeV LHC.
The lower panel shows the N$^3$LO and NNLO predictions normalized to the N$^3$LO prediction.
Taken from \protect\cite{Dulat:2018bfe} (left) and from~\protect\cite{Cieri:2019iiu} (right).}
\label{fig:DWtwo}
\end{figure}

Studies of the properties of the Higgs boson at the 13 TeV LHC are
well under way, and we can look forward to a rich program of precision
exploration of the Higgs sector at the HL-LHC and HE-LHC (see, e.g.,
Ref.~\cite{Cepeda:2019klc,deBlas:2019rxi} for a review), provided
predictions for the relevant observables are well under control.
Given the importance of Higgs production in gluon-gluon fusion via a
heavy quark loop, which is the dominant SM Higgs production mode at
the LHC, significant theory effort went into improving predictions for
both the total rate and kinematic distributions. Recently, two
independent calculations of the Higgs rapidity distribution in $gg \to
H$ production at next-to-next-to-next-to-leading order (N$^3$LO) in
perturbative QCD became
available~\cite{Dulat:2018bfe,Cieri:2019iiu}. They employ very
different methods and their agreement provides a powerful and
important cross-check. In~\cite{Dulat:2018bfe}, an expansion about the
Higgs production threshold is used and the missing terms have been
fixed so that the known, inclusive all-order result is
reproduced. In~\cite{Cieri:2018oms,Cieri:2019iiu} for the first time
the transverse momentum ($q_T$) subtraction formalism is extended to a
N$^3$LO calculation, making use of the fact that N$^3$LO specific
singularities only arise in the $q_T \to 0$ limit which are known
analytically. The impressive reduction of the theoretical uncertainty
from the variation of the renormalization and factorization scale when
including higher-orders in QCD is shown in Fig.~\ref{fig:DWtwo}.

The study of the production of a pair of Higgs bosons is one of the
main goals of the HL-LHC~\cite{Cepeda:2019klc}, since it directly
probes the trilinear Higgs self interaction and thus the shape of the
Higgs potential.  Di-Higgs production in vector boson fusion (VBF),
$pp \to HHjj$, is the second largest di-Higgs production cross section
and is now also available at N$^3$LO QCD~\cite{Dreyer:2018qbw} in the
public code {\sc proVBFHH}. This represents the first N$^3$LO
calculation for a $2\to 4$ process. The calculation of the inclusive
cross section is based on the structure function approach where all
radiation is integrated over. The differential distributions are
obtained by using the projection-to-Born
method~\cite{Cacciari:2015jma} where the inclusive N$^3$LO calculation
is combined with the differential NNLO calculation of di-Higgs
production in association with three jets presented
in~\cite{Dreyer:2018rfu}. The N$^3$LO QCD corrections to differential
distributions studied in~\cite{Dreyer:2018qbw} show a remarkable
stability against scale variations and also against deviations of the
trilinear coupling from the SM value.

While N$^3$LO QCD predictions for the LHC are still few and far
between, state-of-the-art of predictions for SM precision physics at
the LHC are processes with up to two colored particles in the final
state at next-to-next-to-leading order (NNLO) in QCD and $2\to
6$~fermions processes at next-to-leading-order (NLO) in EW theory
(see, e.g., ~\cite{Azzi:2019yne} for a recent review).  For instance,
the {\sc MATRIX} framework~\cite{Grazzini:2017mhc} provides automated
NNLO QCD calculations of fully differential cross sections for the LHC
based on the $q_T$ subtraction formalism. The long list of available
processes includes single electroweak (EW) gauge and Higgs boson
production, di-boson and $t\bar t$ production.
In~\cite{KallweitMoriond} results from a combination of NNLO QCD and
NLO EW predictions for differential distributions in $VV$ ($V=Z,W$)
production processes (including decays into fermion pairs) show the
large reduction of differential cross sections due to NLO EW
corrections, especially at high transverse momenta or invariant
masses. Good theoretical control of di-EW boson (and tri-EW boson)
production especially in these kinematic regions is essential for
precision tests of EW triple (and quartic couplings) thereby searching
for indirect signals of NP. Another recent improvement provided in the
{\sc MATRIX} framework is the consistent combination of the NNLO QCD
contribution to $ZZ$ production in the $q\bar q$ annihilation channel
with the NLO corrections to the loop-induced $gg \to ZZ$
channel~\cite{Grazzini:2018owa}.  This also requires the inclusion of
the $qg$ channel and allows for the construction of an approximate
N$^3$LO prediction.

\begin{figure}
\begin{minipage}{0.66\linewidth}
\centerline{\includegraphics[width=0.7\linewidth]{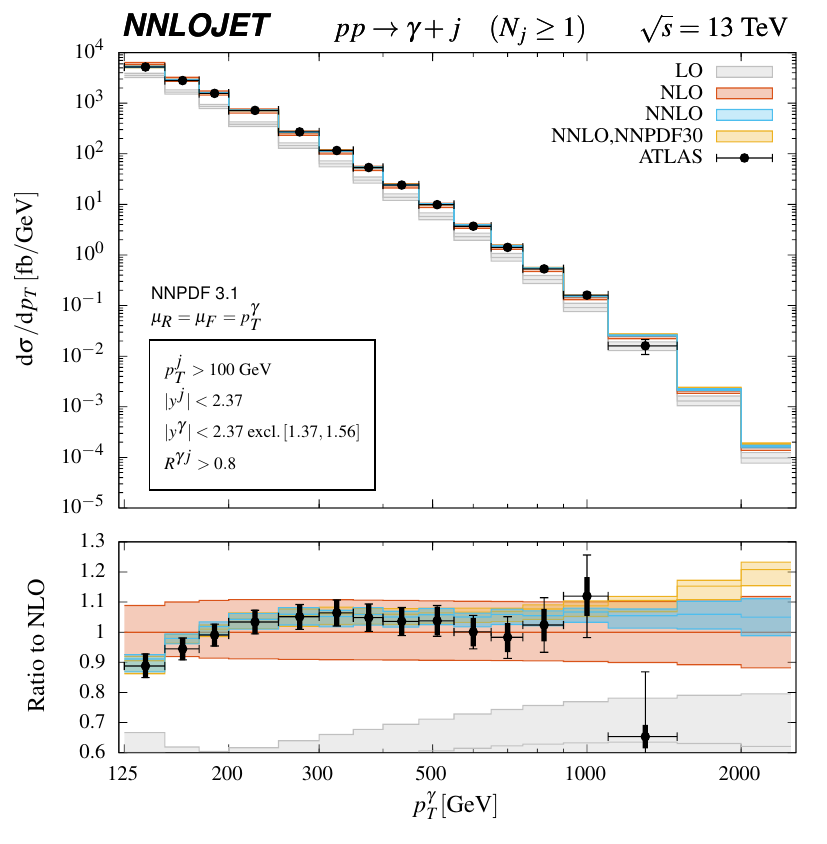}}
\end{minipage}
\caption[]{Predictions for the $p_T^\gamma$ distribution of the photon
  in $\gamma$+jet production, at LO, NLO and NNLO QCD compared to
  ATLAS data. Taken from~\protect\cite{Chen:2019zmr}.}
\label{fig:DWthree}
\end{figure}

An important hadron collider observable which provides a sensitive
test of perturbative QCD and direct access to the gluon distribution
inside the hadron is the transverse momentum of the photon
$p_T^\gamma$ at large $p_T^\gamma$ in isolated photon and photon+jet
production. A new calculation of the $p_T^\gamma$ distribution at NNLO
in QCD with emphasis on studying different prescription for photon
isolation has been presented in~\cite{Chen:2019fil,Chen:2019zmr} (and
compared to ~\cite{Campbell:2016lzl}). Photon isolation criteria need
to be applied to be able to define a cross section which only depends
on the perturbatively calculable direct photon production and not on
the non-perturbative fragmentation of a quark or gluon into a photon
(see, e.g., \cite{Frixione:1998jh}). In~\cite{Chen:2019zmr} a hybrid
approach~\cite{Siegert:2016bre} is found to be closer to the
experimental treatment and to open new ways for perturbative QCD test
of this procedure. Figure~\ref{fig:DWthree} shows the impressive
agreement of the $p_T^\gamma$ distribution at NNLO QCD with ATLAS data
at the few percent level.

To take full advantage of the ever increasing experimental precision
at the LHC, continued advances in performing multi-loop QCD
calculations are needed. For instance, knowledge of the 3-jet
production cross section at NNLO QCD would allow for a precision
measurement of the strong coupling constant $\alpha_s(Q^2)$ at high
energy scale $Q^2$ when extracted from the ratio of 3-jet and 2-jet
rates. One of the challenges in achieving this goal is the analytic
calculation of five-parton scattering amplitudes at 2-loop order. Only
recently the previously unknown massless non-planar master integrals
(MI) became available in terms of Goncharov
polylogarithms~\cite{Goncharov:1998kja} ``thanks to nice
mathematics''~\cite{HennMoriond,Chicherin:2018old}, which completes
the full set of master integrals needed for the analytic calculation
of the five-parton scattering amplitude. How the latter can be
assembled was shown~\cite{ZoiaMoriond} on the example of a ${\cal N}=4$ super-Yang-Mills
theory~\cite{Chicherin:2018yne} and ${\cal N}=8$
supergravity~\cite{Chicherin:2019xeg} (see
also~\cite{Badger:2019djh,Abreu:2019rpt,Abreu:2018aqd}), and is based
on the idea of rational reconstruction~\cite{Abreu:2018jgq}, which
combines 2-loop numerical unitarity results for the MI coefficients
with analytic results for the MI.  These advances in multi-loop
calculations also crucially rely on a systematic understanding of
special functions appearing in Feynman integrals such as the Goncharov
polylogarithms.  As discussed in~\cite{Broedel:2019hyg} in NNLO
calculations for processes involving massive particles in the loops
such as $t \bar t$ production at NNLO QCD or mixed 2-loop QCD-EW
corrections, integrals appear whose analytic calculation requires
elliptic multiple polylogarithms (eMPL).  While MPLs are obtained by
integrating rational functions on a Riemann sphere, eMPLs arise as
integrals on a torus. Much effort is now under way to express these
complex two-loop scattering amplitudes in terms of eMPLs and to find a
formulation of eMPLs suitable for use in these
calculations~\cite{Broedel:2019hyg}.

Apart from improvements in fixed-order calculations, advances are also
needed in the identification and all-order resummation of large
logarithms as well as in the consistent combination of fixed-order and
resumed calculations. Large logarithms appear in processes involving
very different energy scales in certain regions of phase space and may
spoil the perturbative convergence.  The combination is especially
useful for reducing the theory uncertainty due to scale variation for
processes where performing a higher fixed-order calculation is
currently out of reach.  For example, the production of a Higgs boson
in association with a $t\bar t$ pair ($t\bar tH$) is an important SM
Higgs production process, since it allows for direct measurement of
the top-quark Yukawa coupling. Although only recently being discovered
by CMS~\cite{Sirunyan:2018hoz} and ATLAS~\cite{Aaboud:2018urx}, it is
already clear that the NLO QCD prediction for the total cross section
needs to be improved given its large uncertainty due to scale
variation. As shown in~\cite{Kulesza:2019adl}, combining the NLO QCD
prediction for $t\bar tH$ production at the LHC with
threshold-resumed logarithmic contributions from soft gluon emission
at next-to-next-to-leading logarithmic (NNLL) accuracy considerably
reduces the scale uncertainty compared to the NLO result.  It is
interesting to note that the aforementioned {\sc MATRIX} framework
also provides NNLO+NNLL predictions for $VV$
production~\cite{Grazzini:2015wpa}.
 
The current state-of-the-art for resummation (of global logarithms) is
$N^3LL$ accuracy for hadronic inclusive cross sections and event
shapes, but there are observables which exhibit more complicated
radiation patterns resulting in the occurrence of non-global
logarithms~\cite{Dasgupta:2001sh}. Non-global observables are
sensitive to radiation in only part of the phase space, such as the
jet mass discussed in~\cite{Balsiger:2018ezi}.  In Soft-Collinear EFT
(SCET) a factorization theorem for non-global observables allows for
resummation of the non-global logarithms at NLL' accuracy. It is
interesting to note that the renormalization group evolution (RGE) equation 
for the Wilson coefficients in the
SCET formulation is equivalent to a parton shower equation, and a
comparison of the jet observable at NLL+LO with {\sc PYTHIA} is shown
in~\cite{Balsiger:2018ezi}.

Large logarithms can also appear in higher-order BSM predictions for
large NP energy scales, which need to be resumed. For example,
in~\cite{Bahl:2019ccs}, large logarithms appearing in radiative
corrections to the MSSM Higgs masses for $M_{SUSY} \gg m_{top}$ are
resumed up to partial $N^3LL$ accuracy in a SM EFT approach (SUSY
particles are integrated out). The resumed result is then combined
with fixed-order calculations to obtain precise predictions also for
intermediate values of $M_{SUSY}$. Its implementation in {\sc
  FeynHiggs}~\cite{Bahl:2018qog} is then used to define new MSSM Higgs
benchmark scenarios.  In Ref.~\cite{Konig:2019gfm} SCET is used to
describe a NP scenario with a new gauge-singlet heavy spin-0 particle
with mass $M_S$ far above the EW scale $v$, which allows for the
resummation of large logarithms of $M_S/v$ via RGE in predictions for
its decay width to SM particles (see also~\cite{Alte:2019iug}).
  
\section{PDFs, BFKL dynamics, TMD factorization and evolution, and hadronization}

Parton distribution functions (PDF) are an essential component of
predictions for hadron collisions, and PDF uncertainties can be a
limiting factor in high-precision studies of key SM processes and
observables.  For example, the PDF uncertainty in the first $W$ boson
mass measurement at the LHC by ATLAS~\cite{Aaboud:2017svj} is quoted
to be 9~MeV (see, e.~g., \cite{Hussein:2019kqx} for a recent study)
compared to an experimental systematic uncertainty of 11~MeV . It is
therefore of the utmost importance to further improve our knowledge of
the PDFs of the proton, for instance by including new data in global
PDF fits.  Lepton-pair production via the Drell-Yan (DY) process at
the LHC, $pp\to l^+l^-X$ (neutral current (NC)) and $pp\to l\nu X$
(charged current (CC)) ($l=e,\mu$), is an excellent probe of the
structure of the proton. A study in~\cite{Accomando:2018nig} of the
impact of including the forward-backward asymmetry $A_{FB}$ in the NC
DY process at the LHC and HL-LHC on different NNLO PDF sets shows that
indeed $A_{FB}$ has the potential to further constrain the quark
PDF. In~\cite{Accomando:2018nig}, it is also proposed to apply a high
rapidity cut in order to suppress the $d\bar d$-quark luminosity and increase
sensitivity to the up-type (anti)-quark PDFs. This study was done by
using {\sc xFitter}~\cite{Alekhin:2014irh}, which is a framework for
performing PDF fits, studying the impact of including data in the fits,
and a variety of QCD and PDF studies. For instance,
in~\cite{NovikovMoriond} {\sc xFitter} is used to determine the pion
PDF from NA10, E615, and WA70 data.

The DY process also offers the possibility of testing the description of the
high-energy behavior of the scattering of hadrons in QCD~\cite{Golec-Biernat:2018kem}
according to Balitsky, Fadin, Kuraev and Lipatov (BFKL) and the QCD factorization
formula for transverse momentum dependent parton densities
(TMD)~\cite{Balitsky:2017gis}.  In~\cite{Golec-Biernat:2018kem}, the
forward production of a lepton pair in association with a backward jet
is proposed to probe the resummation of large logarithms due to QCD
radiation at high energies via the BFKL
formalism. Especially the angular coefficients $A_i, i=0,1,2$ of the
DY lepton pair offer sensitive tests of the BFKL dynamics in this
process. Moreover, the combination $A_0-A_2$ is sensitive to the TMD
of the gluon~\cite{Golec-Biernat:2018kem}.  Unlike collinear PDFs,
TMDs include non-perturbative information about partonic transverse
momentum and polarization degrees of freedom, and a classic example
for their application is the description of the transverse momentum
$q_T$ distribution of the EW gauge boson in DY production at small
$q_T$.  TMD factorization allows to express differential cross
sections of DY processes at small $q_T$ ($q_T^2 \ll Q^2$, where $Q$ is
the high-mass scale of the hard scattering) as a convolution of the
partonic, hard scattering cross section with TMDs up to large $q_T$
corrections~\cite{Collins:2011zzd} (for $q_T^2 \approx
Q^2$, TMD $\to$ collinear factorization). It is therefore important to
quantify at which $q_T$ these corrections become
important. In~\cite{Balitsky:2017gis} the higher-twist power
corrections to the NC DY process for $s\gg Q^2 \gg q_T^2$ have been
calculated and estimated to be a few percent of the leading twist
result at $q_T \approx Q/4$.  In~\cite{Hautmann:2017fcj} TMDs are
constructed from QCD evolution equations in the Parton Branching (PB)
method at NLO QCD. The PB method has the feature that the splitting
kinematics at each branching can be calculated, similarly to a parton
shower. These TMDs depend on the ordering variable in the branching
and their impact on DY $q_T$ spectra have also been studied in
Ref.~\cite{Hautmann:2017fcj,Martinez:2019mwt}.

Another crucial non-perturbative aspect of hadron collider physics is
the formation of hadrons from quarks and gluons.  Predictions for
hadron production rely on models with tunable parameters implemented
in Monte Carlo event generators (see, e.g. ~\cite{Sjostrand:2016bif}
for a review), such as the Lund string model in {\sc PYTHIA}. In a
simple string or flux tube model $q\bar q$ pairs are created in the
strong color field in the flux tube which then combine to color
singlet hadrons. Ref.~\cite{Koshelkin:2015qna} considers the extension
of a one-dimensional string model which cannot describe transverse
dynamics to a 2-dim. flux tube. It is based on a compactification of
$3+1$-dim. QCD to $1+1$-dim. QCD assuming longitudinal dominance and
transverse confinement. Predictions for $p_T$ and rapidity
distributions in this approach have been derived and compared to NA61
pion production data~\cite{KoshelkinMoriond}. The modeling of baryon
production in the cluster model (see, e.g., \cite{Gieseke:2017clv}),
in particular the role of baryon number in formations of pre-confined
baryonic clusters has been discussed in~\cite{LiMoriond}.

\section{Heavy Ion Collision}

In high-energy heavy ion collisions at the LHC and RHIC, a new form of
matter is produced with unexpected properties. This Quark-Gluon Plasma
(QGP) is understood to be a strongly interacting near perfect
liquid. The study of QGP properties in heavy-ion collisions is a vast
and exciting field of experimental and theoretical exploration, and
for a recent review see, e.~g., Ref.~\cite{Busza:2018rrf}. Here a
brief summary will be given of calculations for jet $p_\perp$ broadening
and Higgs boson production in a QGP, and for some interesting phenomena
observed near $T_c$, i.~e. the temperature where the QCD phase
transition between the confined and QGP phase occurs.

High-energy quarks and gluons traversing the QGP experience a
broadening of the transverse momentum ($p_\perp$) distributions
originating from radiative energy loss in multiple parton scattering
and from QCD radiative corrections.  While Ref.~\cite{Liou:2013qya}
considers the NLO QCD corrections to quark $p_\perp$ broadening in the
soft gluon approximation, Ref.~\cite{Zakharov:2018rst} goes beyond
this approximation and finds large negative contributions not included
in~\cite{Liou:2013qya}. Interestingly this finding seems to be
supported by the recent STAR measurement of the hadron-jet
correlations as stated in~\cite{Zakharov:2018rst}.

Properties of the QGP such as bulk viscosity ($\zeta/s(T)$), speed of
sound $C_s(T)$ and electric conductivity $\sigma_{el}(T)$ are expected
to exhibit interesting behaviors for $T \to T_c$, i.e.  $\zeta/s$
quickly rises, $C_s$ is at its minimum, and $\sigma_{el}$ decreases.
In~\cite{Kerbikov:2018wdl,KerbikovMoriond} it is shown that for quark
matter at moderate density ($\approx 300-400$ MeV) and temperature ($T
\to T_c$ from above, $\sim 20$ MeV) all these phenomena can be traced
back to a common dynamical origin namely the interaction of
phonons(photons) with the soft collective mode of the di-quark
field. This result is derived from the time-dependent Ginzburg-Landau
functional with random Langevin forces.

\begin{figure}
\begin{minipage}{0.66\linewidth}
\centerline{\includegraphics[width=0.7\linewidth]{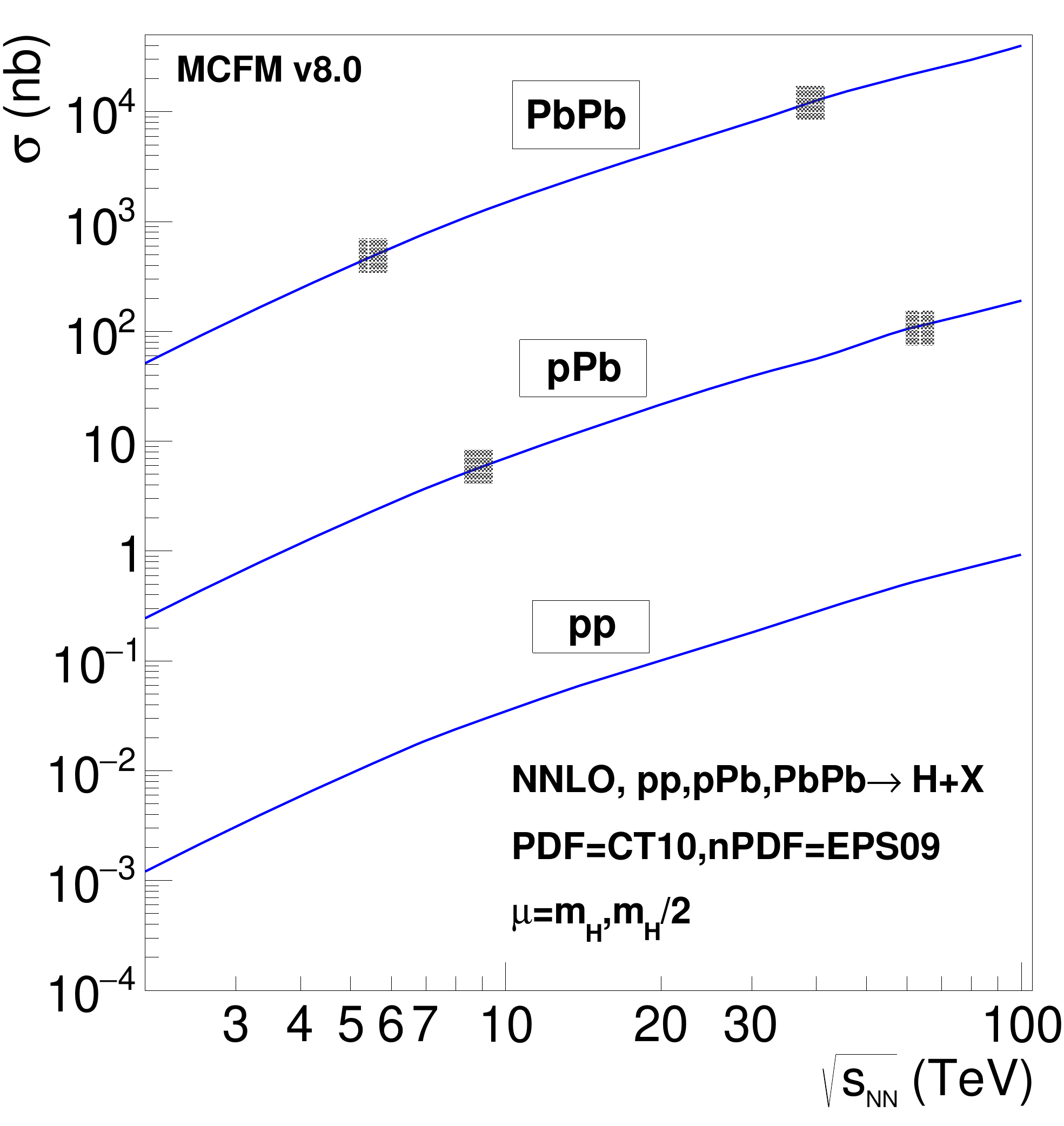}}
\end{minipage}
\caption[]{Total production cross sections for the Higgs boson at NNLO
  in $pp, pPb$ and $PbPb$ collisions as a function of the center-of-mass
  energy $\sqrt{s_{nn}}$ (the shaded boxes indicate the nominal LHC and FCC
  energies).  Taken from \protect\cite{dEnterria:2017jyt}.}
\label{fig:DWfour}
\end{figure}

Finally, the prospect of observing the SM Higgs boson in high-energy
heavy ion collisions is studied
in~\cite{dEnterria:2017jyt,dEnterria:2019jty}, and the enhancements of
the Higgs production cross sections in $PbPb$ and $pPb$ collisions
over the ones in $pp$ collisions at the LHC, HE-LHC and FCC is shown
in Fig.~\ref{fig:DWfour}. It is also interesting to note that the
effect of the medium on the Higgs decay widths to a gluon or light
quark pair only introduces an additional correction of ${\cal
  O}(\alpha_s (T/M_H)^4)$ times the vacuum branching
ratios~\cite{Ghiglieri:2019lzz}.

\section*{Acknowledgments}

I would like to thank the organizers for the invitation, and a perfectly
organized, inspiring conference, featuring excellent presentations
over a wide range of interesting and relevant topics in QCD and
high-energy interactions.  I am grateful to the many speakers, who
patiently answered my questions and provided me with additional
information and explanations. I also gratefully acknowledge support by
the U.S. National Science Foundation under award no.~PHY-1719690.

\section*{References}

\bibliography{refs_moriond2019}

\end{document}